\documentclass[pra,twocolumn,superscriptaddress,aps,showpacs,amsmath,amssymb,nofootinbib]{revtex4}
\usepackage{graphicx}
\usepackage{dcolumn}
\usepackage{bm}
\newcommand{\bce}{\begin{center}}
\newcommand{\ece}{\end{center}}
\newcommand{\beq}{\begin{equation}}
\newcommand{\eeq}{\end{equation}}
\newcommand{\beqa}{\begin{eqnarray}}
\newcommand{\eeqa}{\end{eqnarray}}
\begin{document}

\title{Macroscopic self trapping in BECs: analysis of a dynamical quantum phase transition}

\date{\today}

\author{B. Juli\'a-D\'\i az}
\affiliation{Departament d'Estructura i Constituents de la Mat\`{e}ria,\\
Universitat de Barcelona, 08028 Barcelona, Spain}

\author{D. Dagnino}
\affiliation{Departament d'Estructura i Constituents de la Mat\`{e}ria,\\
Universitat de Barcelona, 08028 Barcelona, Spain}

\author{M. Lewenstein}
\affiliation{ICREA-Instituci\'o Catalana de Recerca i Estudis Avan\c{c}ats, Llu\'{i}s Companys 23,
08010 Barcelona, Spain}
\affiliation{ICFO-Institut de Ci\`{e}ncies Fot\`{o}niques, 08860 Castelldefels (Barcelona), Spain}

\author{J. Martorell}
\affiliation{Departament d'Estructura i Constituents de la Mat\`{e}ria,\\
Universitat de Barcelona, 08028 Barcelona, Spain}

\author{A. Polls}
\affiliation{Departament d'Estructura i Constituents de la Mat\`{e}ria,\\
Universitat de Barcelona, 08028 Barcelona, Spain}

\begin{abstract}
We consider a Bose-Einstein condensate in a double-well 
potential undergoing a dynamical transition from the regime 
of Josephson oscillations to the regime of self-trapping. 
We analyze the statistical properties of the ground 
state (or the highest excited state) of the Hamiltonian 
in these two regimes for attractive (repulsive) interactions. 
We demonstrate that it is impossible to describe the 
transition within the mean-field theory. In contrast, 
the transition proceeds through a strongly correlated 
delocalized state, with large quantum fluctuations, and 
spontaneous breaking of the symmetry. 
\end{abstract}


\maketitle

\section{Introduction}

Cold atom physics, e.g. Bose-Einstein condensates (BEC), 
provides a useful set up to tackle fundamental problems of 
quantum physics at a macroscopic scale. The physics of weakly 
interacting Bose gases has been described very successfully 
by employing a mean field approach, that is the Gross-Pitaevskii 
equation (GPE). Such approach implies that to a certain extent 
quantum fluctuations and correlations are relatively unimportant 
in a large variety of phenomena, see for instance ~\cite{Leggett01}. 

\subsection{Quantum correlations and symmetry breaking}

The mean field approach can be improved by including 
in the theoretical description small (Gaussian) fluctuations 
around the mean field solutions. These fluctuations are described 
then by the Bogoliubov-de Gennes (BdG) equations, and have been 
intensively studied in experiments and theory~\cite{pitaevski2003}. 
The approach based on BdG equations has its limits also. In 
particular, it is completely inadequate in strongly correlated 
quantum systems, such as for instance those that can be realized 
with ultracold gases in optical lattices, or in low dimensional 
gases (for recent reviews see~\cite{bloch-rmp,lewenstein-adv}). 
Interestingly, it has recently been pointed out by several 
authors that the BdG approach breaks down in systems that undergo 
transition between two states that are well described by the 
mean field theory, but that differ dramatically in form, for 
instance having different symmetry properties. Transition 
between such states on the level of mean field description 
is typically associated with dynamical instabilities of the 
GPE, and exponential growth and squeezing of the BdG fluctuation. 
The state of the system in such situations becomes strongly 
correlated and entangled, and its description requires 
necessarily to go beyond the mean field theory. Several authors 
have pointed out this fact in various contexts: sonic analogues 
of ``black holes''~\cite{garay}, vortex nucleation in small 
atomic clouds~\cite{ueda,dagnino09,dagnino09a,wilkin}, BECs 
in rotating ring superlattices~\cite{nunnenkamp}, or in the 
ground state of BEC in a double well potential~\cite{weiss}.

Let us explain how the quantum correlations attain significance 
for the case of vortex nucleation in a rotating BEC~\cite{dagnino09}. 
In this paper exact diagonalization methods for small atomic 
clouds are used and compared with the mean field description, 
indicating directly the impossibility of a correct description of 
the transition within the mean field formalism. By performing 
the exact diagonalization of the Hamiltonian of the system, 
the authors show that in the transition region the atoms are 
far from being in the same single particle state. Instead, 
two macro-occupied single particle states (MSPS) appear 
with comparable occupations. 

In this article we will show how the physics of BECs in a 
double-well resembles the above scenario. In the case of the 
rotating BEC, the control parameter driving the transition 
is the rotation frequency. In the case of the BEC in a 
double-well, it will be the interaction between the atoms, 
or, correspondingly, the height of the potential barrier 
splitting the BEC. Contrary to the previous works, that 
focus on the properties of the ground state, or low energy 
states of the considered system, we deal here with the 
dynamical transition both for attractive and repulsive 
interactions, for which also the properties of the highest 
energy, or high energy states are relevant. 

\subsection{BECs in double well potentials} 

The physics of BECs on a double-well has attracted a 
great deal of attention since the first theoretical 
studies predicting the presence not only of Josephson-like 
oscillations of population, but also of self-trapped 
states, see Refs.~\cite{Smerzi97,Milburn97}. Its recent 
experimental characterization by the Heidelberg 
group~\cite{Albiez05} opens the possibility of practical 
applications of these bosonic junctions. 

As pointed out in Ref.~\cite{Leggett01}, most of the 
physics of bosonic Josephson junctions, including self-trapping, 
$\pi-$modes and Josephson oscillations, can be understood 
semiclassically. We 
will describe, however, how the transition from the 
Josephson-regime to the self-trapped regime is indeed 
dominated by quantum correlations. To that extent we 
will consider the exact diagonalization of the Bose-Hubbard 
(BH) Hamiltonian, and compare the results always to 
predictions of the semiclassical approach (in which 
quantum mechanical operators are replaced by their 
$c$-number averages). 

Our aim is to isolate the purely quantum aspects of the 
transition, defined as those which are either not 
understandable in terms of a mean field description of 
the problem or, alternatively, by an elementary semiclassical 
analysis.

\subsection{The plan of the paper}

We will start, sections II-IV, discussing the properties 
of the BH Hamiltonian emphasizing the relevance of the 
condensed fractions, eigenvalues of the one-body density 
matrix, and the symmetry properties of the eigenstates 
of the Hamiltonian. By analyzing several quantities 
simultaneously we are able to classify the different 
regimes into which the transition is divided and which 
are clearly beyond the mean field approximation taken 
previously by some authors. The quantum nature of the 
transition is clearly visualized. Then, in section V, we 
present the  dynamical consequences of the 
statical properties described before. The final 
section, VI, gives the conclusions of our analysis.

\section{Description of the model}

As described in ~\cite{Milburn97,Leggett01} the two-mode Bose-Hubbard 
Hamiltonian for $N$ atoms in the simple case of a double-well may be 
written as, 
\beqa
H&=&{-U\over2} \left(\hat{n}_L (\hat{n}_L-1)+ \hat{n}_R(\hat{n}_R-1)\right) 
\nonumber \\
&-& J \left(a^\dagger_Ra_L + a^\dagger_La_R\right) 
- \epsilon (\hat{n}_L-\hat{n}_R)
\eeqa
where $a^\dagger_{L} |n_L,n_R\rangle = \sqrt{n_{L}+1}|n_L+1,n_R\rangle$, 
$a_{L} |n_L,n_R\rangle = \sqrt{n_{L}}|n_L-1,n_R\rangle$, and 
$[a_i^\dagger,a_j]=\delta_{i,j}\,,\quad i,j=L,R$. $L(R)$ stands for the 
left (right) side of the well. Our sign convention is such that 
$U>0 (U<0)$ corresponds to attractive (repulsive) atom-atom interactions. 
We discuss both cases simultaneously as the technicalities are 
similar for both cases. We take $J>0$. 

A small bias, $0<\epsilon\ll J$, is introduced which will ensure 
breaking of left-right symmetry. Numerically we consider, 
$\epsilon/J=10^{-10}$ in all the results presented, which 
intents to provide a realistic implementation of the 
symmetric double well. 

The Hamiltonian is diagonalized in the $N+1$ dimensional space 
spanned by the basis: $\{ |N,0\rangle, |N-1,1\rangle, \dots$, 
$|1,N-1\rangle, |0,N\rangle \}$. With, $|n_L,n_R\rangle = 1/\sqrt{ n_L! n_R!}(a^\dagger_L)^{n_L} 
(a^\dagger_R)^{n_R} |{\rm vac}\rangle$. The dynamics of the system is 
governed by the ratio, $\Lambda\equiv NU/J$, which is controlled 
either by varying the atom-atom interactions $U$, or, by changing the 
barrier height, $J$. 

Any single particle state may be written as, 
$|\Psi_1 (\theta,\phi) \rangle=\cos(\theta/2) |L\rangle
+e^{\imath \phi}\sin(\theta/2)|R\rangle$, a representation 
usually employed to describe single qubits~\cite{libronc}. We define, 
$|L(R)\rangle \equiv a^\dagger_{L(R)}|{\rm vac}\rangle$, and 
$|\pm> \equiv (1/\sqrt{2})(|L\rangle \pm |R\rangle)$. 

The most general $N$-body state can be written as, 
$|\Psi\rangle=\sum_{k=1,N+1} c(k) |N+1-k,k-1\rangle$. The 
number of atoms in each well for a given state is, 
$N_\beta= \langle \Psi| a^\dagger_\beta a_\beta |\Psi\rangle$, with 
$\beta=L,R$. The population imbalance of a state $\Psi$ is defined 
as, $z=(N_L-N_R)/N$. The time evolution of a state $\Psi(0)$ is 
governed by the time-dependent Sch\"rodinger equation.

To characterize the degree of condensation of the system at 
any given time we will make use of the one-body density 
matrix, $\rho$. For a state, $\Psi$, we have, 
$
\rho_{ij}= \langle \Psi | \hat{\rho}_{ij}| \Psi \rangle\,, 
$
with $\hat{\rho}_{ij}= a^\dagger_i a_j\,,
$
and $i,j=L,R$. The trace of $\rho$ is normalized to the total number 
of atoms, $N$.  We will denote its two eigenvectors as $\psi_{1(2)}$ 
and its normalized eigenvalues (divided by the total number 
of atoms $N$) as $n_{1(2)}$, with $n_1\geq n_2\geq 0$. We always 
have, $n_1+n_2=1$.  $n_i$ will correspond to the condensed 
fraction in the macro-occupied state $\psi_i$~\footnote{Other 
authors, e.g. Ref.~\cite{jame05,jame06}, introduce the operators, 
$J_x=(1/2)(a^\dagger_La_R+a^\dagger_Ra_L)$, 
$J_y=(1/2\imath)(a^\dagger_La_R-a^\dagger_Ra_L)$, and 
$J_z=(1/2)(a^\dagger_La_L-a^\dagger_Ra_R)$, which are related to 
$\rho$ though, $\rho_{LL}=N/2 +\langle J_z\rangle$, 
$\rho_{RR}=N/2 -\langle J_z\rangle$, 
$\rho_{LR}=\langle J_x+ \imath J_y \rangle $, and $\rho_{RL}=\langle J_x -
\imath J_y \rangle$.}.

A measure of the spread of the state in the Fock basis will 
be given by the function $S=-\sum_{k=1}^{N+1} |c(k)|^2 \log (|c(k)|^2)$. 
$S$ is positive definite and has a maximum value for the equally 
populated state, $c^2(k)=1/(N+1)$, and is zero for a maximally 
localized state, $c(k)=\delta_{k,k_0}$. 

To clearly identify the region where the quantum fluctuations 
will play a major role we will always compare our numerical 
results to semiclassical predictions. The semiclassics of 
the problem may be obtained by replacing the creation and 
anhilation operators by $c$-numbers,  
$a_{L(R)}=\sqrt{n_{L(R)}}e^{\imath \varphi_{L(R)}}$. We can define 
the phase difference as, $\varphi=\varphi_R-\varphi_L$. The 
semiclassical Hamiltonian reads, 
\beq
{H\over NJ}=-\sqrt{1-z^2}\cos \varphi -{\Lambda\over 4N} (N z^2+N-2)\,,
\eeq
and the time evolution of the population imbalance and 
phase difference is obtained through the Heisenberg 
evolution equations, $i\dot{a}_L=-[H,a_L]$ and $i\dot{a}_R=-[H,a_R]$. 
The resulting equations are, 
$
{\dot{z}(t)\over 2J}=- \sqrt{1-z^2} \sin\varphi
$, and 
${\dot{\varphi}(t)\over 2J}= -{\Lambda\over2} z +{z\over
  \sqrt{1-z^2}}\cos\varphi\,,$ which were originally obtained 
by Smerzi {\it et al.}~\cite{Smerzi97} as the two-mode 
approximation to the Gross-Pitaevskii mean field description 
of the problem. 

Ref.~\cite{Smerzi97} defined the transition from the Josephson 
to the self-trapped regime as a dynamical feature, that is, a 
state initially prepared with a certain $(z(0),\varphi(0))$ 
would undergo Josephson oscillations or would remain self-trapped 
on one side of the double-well if $\Lambda$ is  
larger than a critical value. We will link that definition, which 
is a semiclassical one in our model, with the static properties 
of the BH Hamiltonian. 

\section{General properties of the spectrum}

Following the discussions of Refs.~\cite{jame05}, let us first present 
the spectral decomposition, $|c_\lambda(k)|^2$, of the eigenvectors 
of the Hamiltonian, $H\Psi_\lambda=E_\lambda \Psi_\lambda$, as a 
function of $\Lambda$, see Fig.~\ref{figa1}. Fig.~\ref{figa2} 
concentrates on the properties of the ground state and the highest 
excited state of the spectrum as we vary the value of $\Lambda=NU/J$. 
The main features relevant for our discussion are:

\begin{figure}[t]
\includegraphics[width=0.4\columnwidth, angle=0, clip=true]{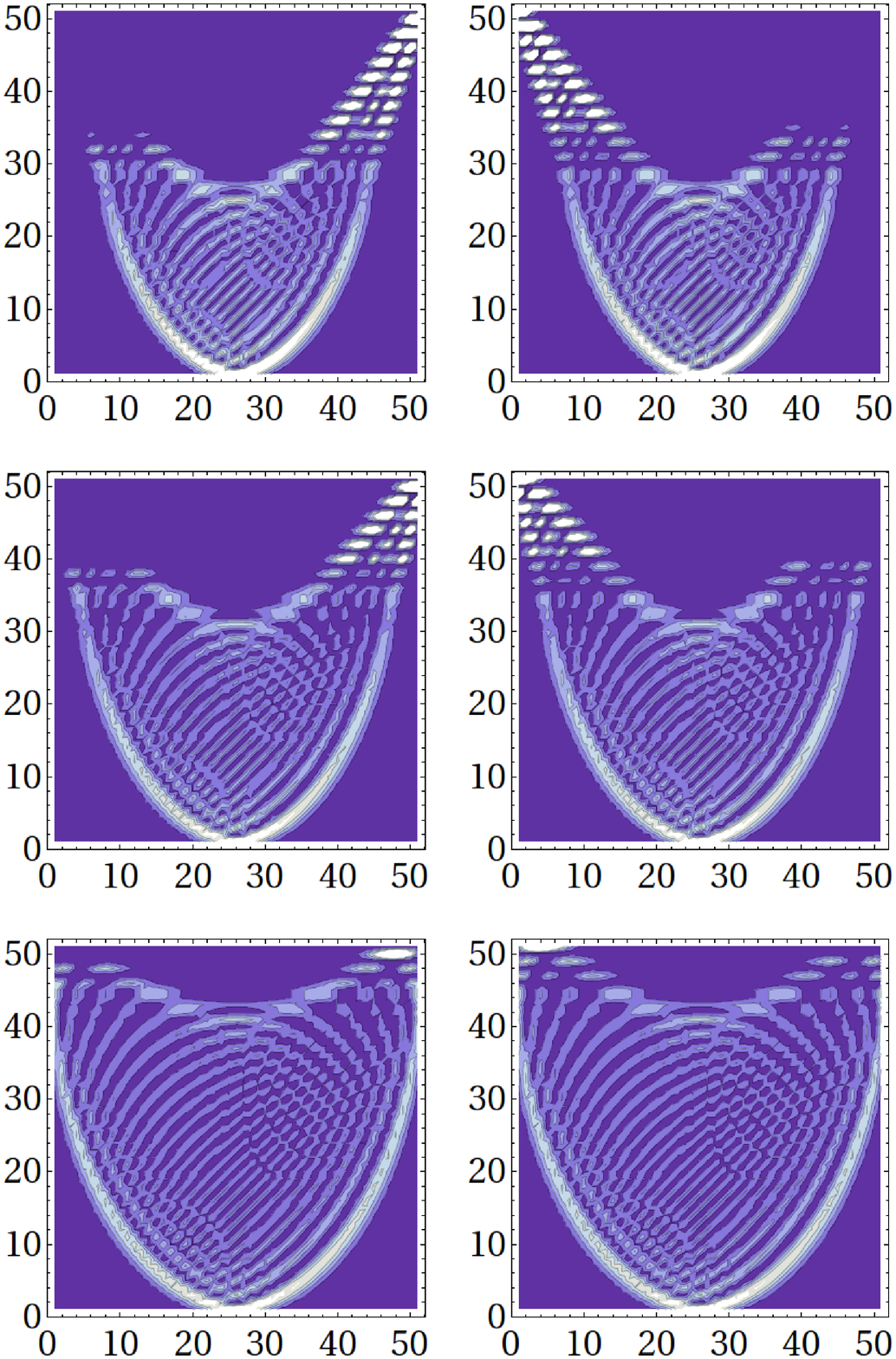}
\hspace{1cm}
\includegraphics[width=0.4\columnwidth, angle=0, clip=true]{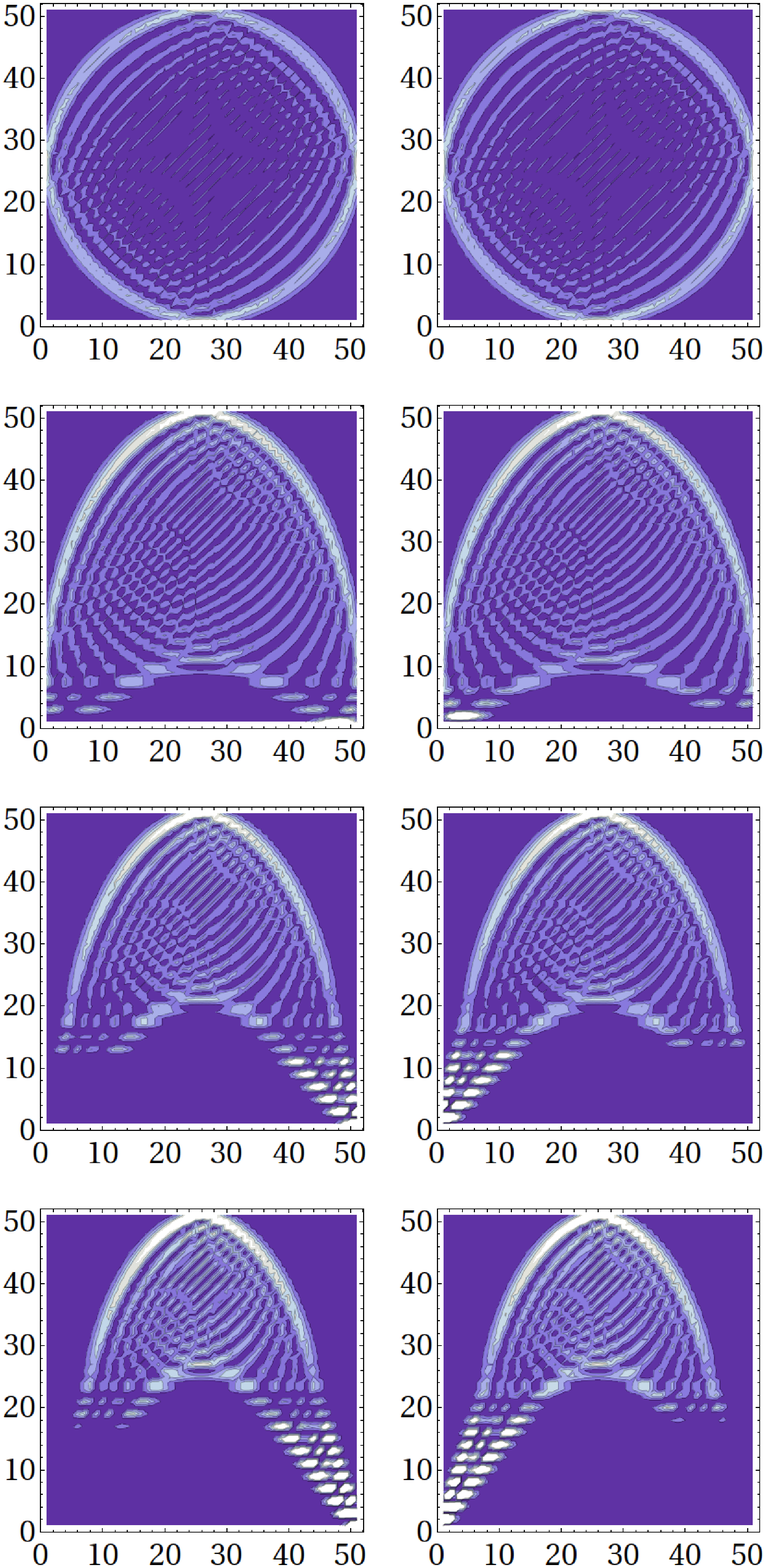}
\caption[]{Density plot of $|c_\lambda(k)|^2$, the horizontal (vertical) axes 
run through $k=1,\dots, N+1$ ($\lambda=1,\dots,N+1$), for, starting 
from above, (left) $NU/J=-4,-8,-12$ and (right) $NU/J=0, 4, 8, 12$. The left (right) 
column of each two displays the eigenvectors with positive (negative) 
imbalance. Blue corresponds to zero and white to the maximum 
value. $N$ is $50$.}
\label{figa1}
\end{figure}
\subsection{Attractive interactions: ground state analysis}

\begin{figure}[t]
\includegraphics[width=0.9\columnwidth, angle=0, clip=true]{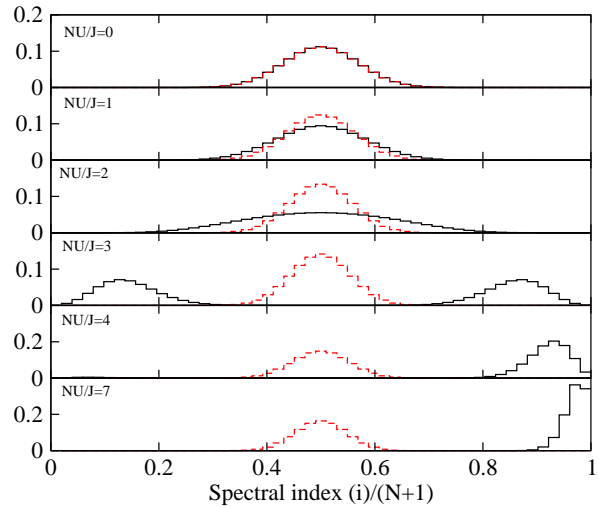}
\caption[]{Spectral decomposition, $|c_\lambda(k)|^2$ of the 
ground state (solid, black) and the highest state of the Hamiltonian 
(dashed, red) in the basis spanned by $(n_L,n_R)=\{(N,0), (N-1,1),\dots,(0,N)\}$ 
for $N=50$. If $U \to -U$ the same plots would be obtained, but 
with the roles of the ground and highest excited states exchanged.}
\label{figa2}
\end{figure}

When $\epsilon=0$ 
the ground state 
is symmetric and binomial for the 
non-interacting case, 
$\Psi_{GS}= (1/ \sqrt{N!}) [(1/\sqrt{2})(a^\dagger_L+a^\dagger_R) ]^N |{\rm
  vac}\rangle$. 
Then, as the interaction is increased the ground state becomes wider. 
Then, it becomes degenerate with the next excited state, and its 
distribution has eventually two differentiated peaks, thus becoming 
cat-like~\cite{cirac98}. When $\epsilon>0$, but small, Figs.~\ref{figa1}, 
and~\ref{figa2} show that, as we 
increase the interaction 
further, each of the formerly degenerate pairs develops a certain population 
imbalance and the distribution becomes peaked on only one region 
of the Fock space. It is important to note that the left-right 
symmetry is broken for a certain value of $\Lambda$,  thus, the 
ground state will spontaneously acquire a large imbalance.
Our figure differs from Fig. 2 of Ref~\cite{jame05}, because 
we have artificially broken the symmetry with the help of a 
small bias. In this way the two degenerate eigenstates are distinguished 
according to their imbalance. Our figure emphasizes the symmetry breaking 
pattern, which occurs when the system evolves from cat-like to self-trapped. 

\subsection{Repulsive interactions: analysis of the highest energy state}

The case of repulsive interactions is similar but the role 
played by the ground state is now played by the highest 
excited state of the Hamiltonian. With no interaction, the 
highest eigenvector of the Hamiltonian is also binomial, 
$\Psi_{\rm HE}= {1/\sqrt{N!}}[(1/\sqrt{2})(a^+_L-a^+_R) ]^N |{\rm vac}\rangle$, 
as the interaction increases, undergoes similar features as the ground 
state did in the attractive case. The ground state in this case, 
on the contrary, remains mostly binomial with a certain 
squeezing, as also discussed in Ref.~\cite{java99}.

\begin{figure}[htb]
\includegraphics[width=0.9\columnwidth, angle=0, clip=true]{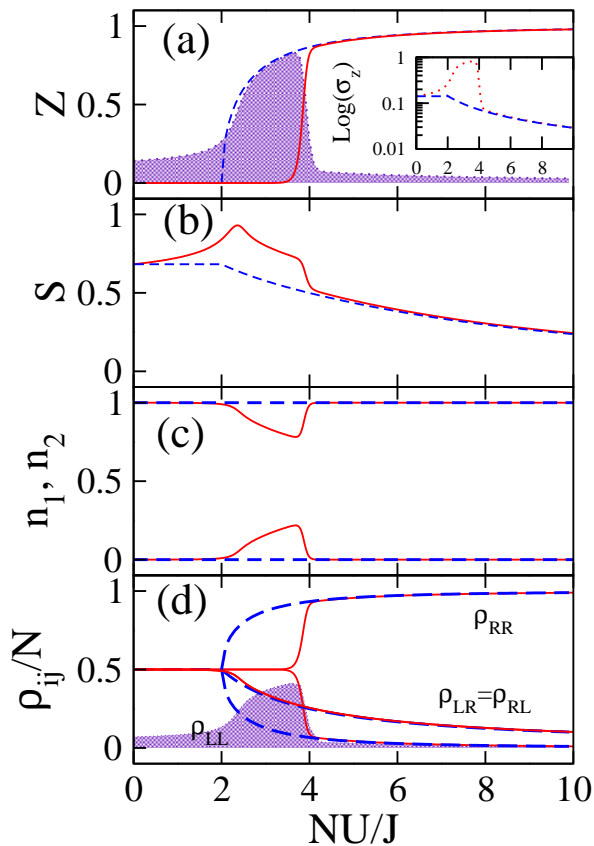}
\caption[]{ (a) Solid line: population imbalance $z$ of the ground state.  
The shaded region corresponds to its dispersion, 
$\sigma_z$. The inset provides 
a logarithmic view of $\sigma_z$, computed from the BH (dotted) and 
its semiclassical approximation (dashed).  
(b) A measure of the spread of the ground state, $S$, normalized 
to its maximum value, which would correspond to a equidistributed state.
(c) Solid and dotted lines depict the condensed 
fractions, $n_1$ and $n_2$, of the one body density matrix of the 
ground state of the Hamiltonian, with imbalance $\geq 0$ as a 
function of $\Lambda=N U/J$. 
(d) The four matrix elements of $\rho$ are depicted as a 
function of $N U/J$. The shaded region corresponds to 
the variance of the diagonal elements, $\sigma_{LL}$, where 
$\sigma_{ij}\equiv \sqrt{|\langle \hat{\rho}_{ij}^2\rangle 
-\langle \hat{\rho}_{ij}\rangle^2|}$ ($\sigma_{RR}$ turns out 
to be equal to $\sigma_{LL}$). The variance of the off-diagonal 
ones is negligible in this plot. The dashed lines in all the 
plots correspond to the semiclassical predictions (at $\epsilon=0$): 
$z_{\rm s.c.}=0$ if $\Lambda<2$, 
$z_{\rm s.c.}=\sqrt{1-4/\Lambda^2}$ if $\Lambda\geq 2$. 
$\rho_{L,L}^{\rm s.c.}=N/2+N z_{s.c.}/2$, 
$\rho_{R,R}^{\rm s.c.}=N/2-N z_{s.c.}/2$, 
$\rho_{L,R}^{\rm s.c.}=\rho_{R,L}^{\rm s.c.}=
N(\sqrt{1-z_{\rm s.c.}^2})/2$. The number of atoms is, $N=50$. }
\label{figa3}
\end{figure}

\section{Attractive interactions: Quantum fluctuations}

To characterize the problem more deeply, we now turn to other 
static properties of the eigenstates. In Fig.~\ref{figa3} we 
present several properties of the ground state as we increase the 
interaction, for the case of attractive interactions. The 
first magnitude is the population imbalance. As seen 
from the figure, it remains zero until a certain value (4 
for $N=50$) of $\Lambda$,
then it grows abruptly and approaches 
1 as $\Lambda$ increases. The semiclassical approximation
would predict 
such behavior to occur at $\Lambda=2$. 
The discrepancy between 
the semiclassical prediction and the observed quantum behavior is 
diminished as the number of atoms, $N$, is increased. The 
figure also shows a shaded region corresponding to 
$\sigma_z\equiv \sqrt{\langle z^2 \rangle  - \langle z \rangle^2}$. 
In the region where the semiclassics fails, the dispersion of 
$z$ becomes large. This agrees with the previous discussion of 
Figs.~\ref{figa1},and~\ref{figa2}, and corresponds to the 
region where the ground state gets wider and, eventually, 
cat-like~\footnote{Employing the language of other authors, 
e.g. Ref.~\cite{jame05}, we would have that for such region 
the semiclassical assumption does not hold, 
$\langle \{J_z,J_z\} \rangle \neq 2 \langle J_z\rangle 
\langle J_z\rangle$.}. The authors of Ref.~\cite{tripp08} consider 
the population imbalance as a suitable order parameter to 
characterize the transition.

The one body density matrix turns out to be a good 
indicator of where the semiclassics fails to describe the 
full quantum results. In the third panel we present the 
condensed fractions, $n_1$ and $n_2$, for different values 
of $\Lambda$. In the semiclassical limit, the system remains 
always fully condensed regardless of the value of $\Lambda$. 
This is clearly not the case in the transition region. 
For $N=50$, the exact dynamics differs in the region $2 \lesssim \Lambda \lesssim4$. 
In that region there are two MSPS, thus being impossible to 
describe the system 
within a mean field formalism. The macro-occupation can be 
traced back to the four elements of the one body density 
matrix, see panel (d). The off-diagonal ones are well described 
by the semiclassics in the whole considered domain, but the diagonal 
ones remain constant (=1/2) for  $\Lambda \lesssim 4$. The dispersion 
of the diagonal matrix elements is directly related to 
$\sigma_z$, and is again large in the transition region. 
The dispersion of the off-diagonal elements is always of 
the order of $1\%$ (thus explaining the agreement with the 
semiclassical picture). 

The large quantum fluctuations seen in $z$, or $\rho_{ii}$, cannot 
be described in a mean field description. The inset in panel (a) 
of Fig.~\ref{figa3} shows $\sigma_z$ computed assuming the state 
of the system corresponds to a mean-field state 
$\Psi_{\rm MF} = [ |\Psi_1(\theta,\phi) \rangle ]^{\otimes N}$, with  
$\theta, \phi$ taken from their semiclassical values. As shown, 
both the full quantum result and the mean-field one 
agree at $\Lambda=0$ and for $\Lambda\gtrsim 4$, but strongly 
disagree in the transition region. Finally, the spread of the state 
in the Fock basis, $S$, is presented in (b). The function has its 
maximum spread for $\Lambda\sim 2.4$, then falls and has an 
abrupt fall off when the ground state evolves from cat-like to 
self-trapped, $\Lambda \sim 4$. 

\begin{figure}[tb]
\includegraphics[width=0.9\columnwidth, angle=0, clip=true]{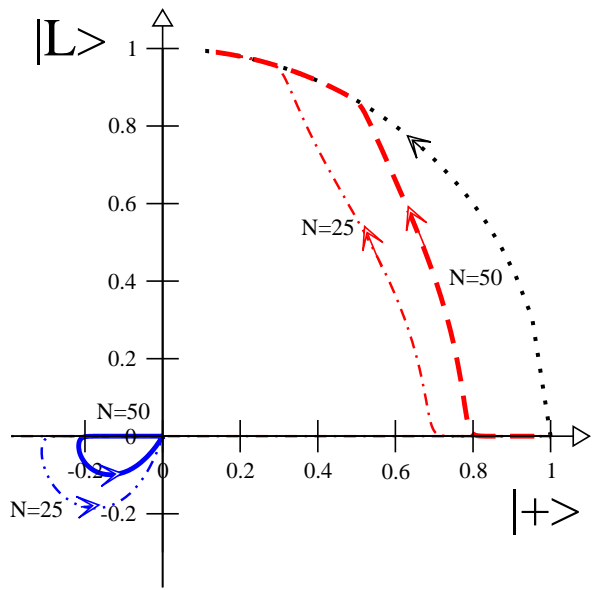}
\caption[]{Evolution of the MSPS of the ground state of the system 
as a function of $\Lambda$. The dashed and solid lines depict the 
BH calculation and correspond to $\psi_1$ and $\psi_2$, respectively. 
The MSPS are represented as 
$|\psi\rangle= \cos (\theta/2) |L\rangle + \sin(\theta/2) |R\rangle$. 
Their modulus is set to the condensed fraction of the state, $n_i$.  
The dotted line corresponds to the semiclassical prediction. 
All lines are obtained by evolving from $\Lambda=0$, to $15$. 
BH results are given for $N=25$ and $N=50$. (This figure 
corresponds to a vertical cut of the Bloch sphere used in 
Fig.~\ref{figa6}.)}
\label{figa4}
\end{figure}

The corresponding evolution of the two MSPS is given in fig.~\ref{figa4}. 
There we plot in the ($|L\rangle$, $|R\rangle$), ($|+\rangle, |-\rangle$) 
plane their evolution as we increase the value of 
$\Lambda$. The exact results behave as follows: for $0 \lesssim \Lambda
\lesssim 2$ 
the ground state of the system is fully condensed on the 
$|+\rangle$ state. It is thus symmetric, $z=0$. For $2<\Lambda \lesssim 3.5$ 
the same two single particle states are macro-occupied in the 
entire domain, see Fig.~\ref{figa3}: $\psi_1=|+\rangle$, and 
$\psi_2=|-\rangle$. The system remains symmetric, $z=0$. Their 
condensed fractions vary with $\Lambda$. For $3.5\lesssim \Lambda \lesssim 4.5$ 
the system has still two MSPS but which depart from the 
$|\pm\rangle$ axis. These MSPS change continuously as we vary 
$\Lambda$, acquiring a non-zero $z$. Finally, for $\Lambda \gtrsim 4.5$ 
the system is fully condensed again, and $\psi_1$ approaches 
$|L\rangle$ as we increase $\Lambda$. The figure is constructed 
in such way that the departure from the semiclassical description 
becomes apparent immediately: any point outside of the circumference 
of radius one is beyond that appoximation. The dynamics is 
therefore mean-field like, both in the extreme Rabi regime, 
$NU/J\to 0$, and in the self-trapped regime, $NU/J \gtrsim 4.5$ for 
$N=50$. The transition region cannot be described within a 
mean-field theory. The semiclassical approximation 
fails to describe the observed behavior almost in the entire 
domain of the transition. 

The dependence of the described static properties as we vary 
$N$ can be summarized as follows: increasing the number of 
atoms, the transition region gets reduced and thus the 
agreement with the semiclassical results, which predict 
a mean-field picture, is improved. As an example, 
Fig.~\ref{figa4} shows also the case of $N=25$. 

The dependence on the bias, $\epsilon$ is qualitatively similar: as 
$\epsilon$ is increased, the jump shifts to smaller values of 
$\Lambda$. In possible experimental set-ups it is however 
important to consider, as done in this work, $\epsilon$ which 
are almost negligible to ensure the transition region 
is broad enough in $\Lambda$. 

\section{Dynamics: The ground and the highest energy state}

Up to now we have concentrated on describing the different 
phenomena occurring near the transition region by analyzing 
the static properties of the spectrum of the Hamiltonian 
as we varied the parameter $\Lambda$. Now we will consider 
the  dynamical consequences. To this extent, we analyze 
the dynamics of the system with a fixed number of atoms, 
$N$, with a maximal initial population imbalance, $z(0)=1$. 

\begin{figure}[tb]
\includegraphics[width=0.9\columnwidth, angle=0, clip=true]{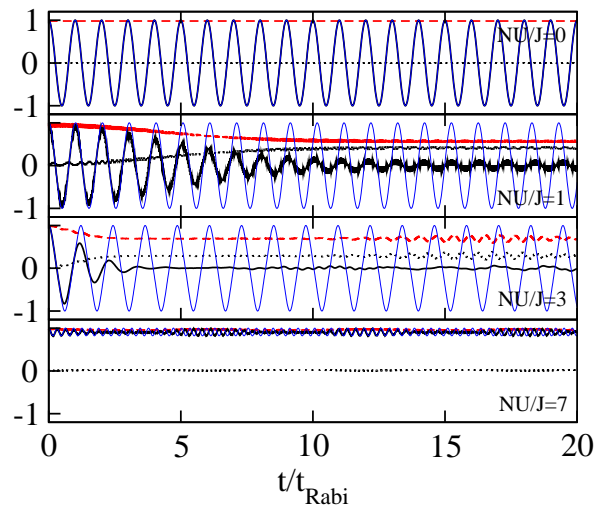}
\caption[]{Evolution of the population imbalance, thick-solid, and of 
the two macro-occupied condensed fractions, $n_1$ and $n_2$, dashed 
and dotted lines, as a  function of time for a state initially prepared 
with $z(0)=1$, $\Psi(0)=|N,0\rangle$. The thin solid line corresponds
to the semiclassical calculation of the population imbalance. The 
semiclassical values of $n_1$ and $n_2$ are the same for all panels 
and are constant and equal to 1 and 0, respectively. The panels 
correspond to different values of $NU/J$, in all cases with $N=50$. }
\label{figa5}
\end{figure}

The simplest quantity which shows already quantum fluctuations 
is the population imbalance. In figure~\ref{figa5} the discrepancies 
between the semiclassical result and the exact solutions for the 
time evolution of the BH model are easily spotted. The BH model, 
with finite $N$, always brings in some damping to the oscillations, 
followed by revivals, in the Josephson regime~\cite{Milburn97}. The 
semiclassical results are seen to be accurate both for the case of 
no interaction among the atoms, Rabi oscillations, or the most 
self-trapped case. In the same figure we depict also the condensed 
fractions, $n_1$ and $n_2$. The semiclassical prediction for 
these quantities is always $n_1(t)=1$ and $n_2(t)=0$. In the finite 
$N$ case however, we see how the condensed fractions do depart from 
1 and 0, specially for the case $1<NU/J \lesssim 4$.

\begin{figure}[t]
\includegraphics[width=0.49\columnwidth, angle=0, clip=true]{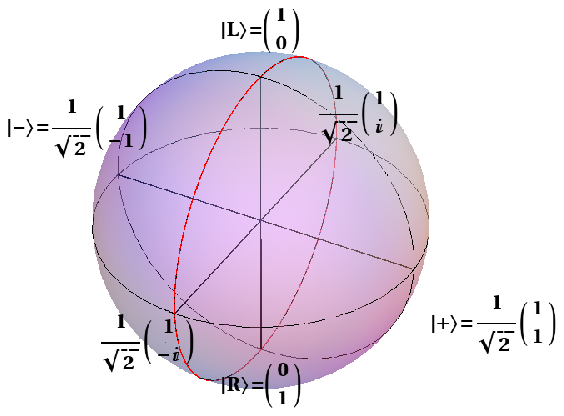}
\includegraphics[width=0.49\columnwidth, angle=0, clip=true]{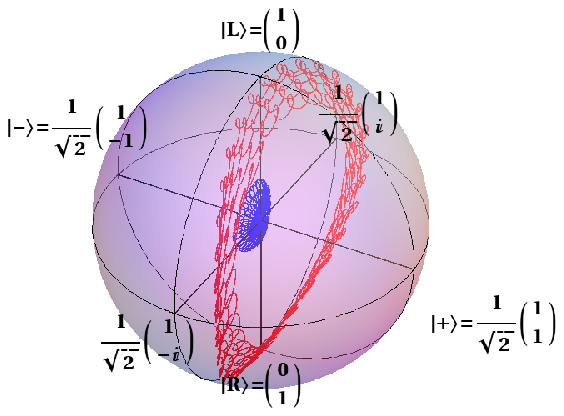}
\includegraphics[width=0.49\columnwidth, angle=0, clip=true]{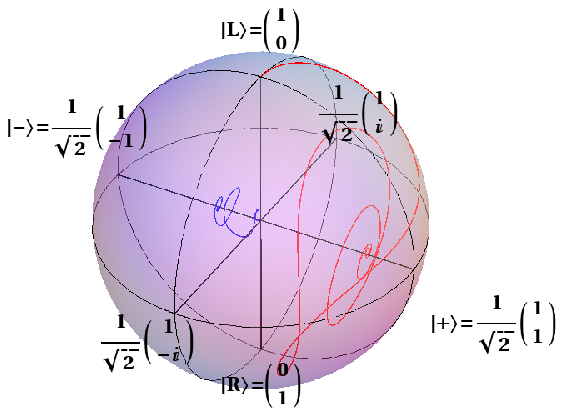}
\includegraphics[width=0.49\columnwidth, angle=0, clip=true]{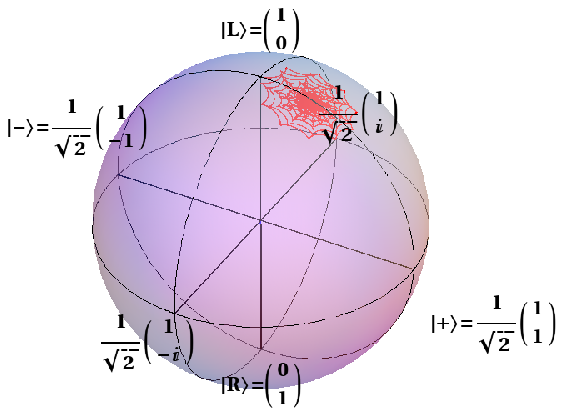}
\caption[]{3D representation of the eigenvectors of the 
one body density matrix, $\psi_{1(2)}$ for $NU/J=0,1,3,7$
as a function of time. The total time is $5 t_{\rm Rabi}$. 
The points are represented as follows: $\theta$ and $\phi$ 
are obtained from, $|\psi_{1(2)} \rangle= 
\cos(\theta/2) |L>+e^{\imath \phi}\sin(\theta/2)|R\rangle$. The 
distance to the origin is equal to $n_{1(2)}$. The initial 
state is always $z=1$ with complete condensation on the 
$|L\rangle$ state. We mark also the $|\pm\rangle$ single 
particle states.}
\label{figa6}
\end{figure}

In the case of attractive interactions the transition from Josephson 
to the self-trapped regime is completely related to the properties of 
the ground state of the system. Essentially, an initial state prepared 
with large imbalance and with $\Lambda$ larger than a critical value 
will remain trapped due to its large overlap with the ground 
state of the system. Correspondingly, the self trapping for the case 
of repulsive interactions is directly related to the highest 
excited state of the Hamiltonian, which is the one with larger 
imbalance in this case. 

In fig~\ref{figa6} we depict the time evolution of the 
two eigenvectors of $\rho$. The length of the eigenvector, 
$\psi_i$, in the Bloch sphere is set to its condensed fraction, 
$n_i$. Therefore, a fully condensed evolution will remain always 
on the surface of the sphere of radius 1. All other situations, 
involving two macroscopically occupied states, would fall 
inside the sphere.

The non-interacting case, $\Lambda=0$, sets the Rabi oscillation 
time, $t_{\rm Rabi}=\pi /J$. The population imbalance performs 
periodic oscillations of maximum amplitude with the corresponding 
frequency, $\omega_{\rm Rabi}= 2 J$. The system remains always condensed 
on a single particle state, see Fig.~\ref{figa6}. 

A small interaction, $NU/J=1$, already changes the picture. First, 
in this case the system is no longer condensed at all times. Now, 
see second panel of fig~\ref{figa5} and the corresponding in 
fig.~\ref{figa6}, as time increases the highest condensed 
fraction, $n_1$, goes from 1 to $\sim 0.6$ after 
$t>10 t_{\rm Rabi}$. The two MSPS rotate 
around the $(|+\rangle, |-\rangle)$ axis, and slowly approach the 
$|+\rangle$ and $|-\rangle$ states with a certain condensed fraction. 

For $NU/J=4$ self-trapping starts.
The condensed fraction is large, 
$\sim 0.9$, but there are sizeable fluctuations. The behavior of 
the MSPS is now different, in this case they do not 
approach the $|+\rangle$ axis, but instead keep rotating around 
an axis closer to the $|L\rangle$ vector. 

Larger interactions make the system more self-trapped and also 
to remain mostly condensed on a state progressively closer to 
$|L\rangle$. The size of the fluctuations seen in the condensed 
fractions decrease as we increase $NU/J$. 

\section{Conclusions}

We have scrutinized the transition from the Josephson to the 
self-trapped regime in BECs, for both attractive and 
repulsive atom-atom interactions. First, we have demonstrated 
the impossibility of a mean-field description of the quantum 
transition for a finite number of atoms, $N$, for attractive 
interactions. The nature of the transition, which involves 
the spontaneous breaking of the left-right symmetry, is 
governed by large quantum fluctuations 
which are not captured by a mean-field description of the 
problem. The ground state of the system in the transition is 
built of two macro-occupied single particle states. Both 
for attractive and repulsive interactions we have shown 
how the self-trapping regime is related to the existence of 
imbalanced eigenstates in the spectrum. 

An extremely challenging experimental proposal emanates naturally 
from this article, namely, the full characterization of a quantum 
phase transition in a BEC, either by considering a varying barrier 
height in the double-well or by modifying the scattering length of 
the atom-atom interaction. 

B.J-D. is supported by a CPAN CSD 2007-0042 contract. The 
authors thank D. Sprung for a careful reading of the manuscript 
and N. Barber\'an for discussions at early stages of the project. 
This work is also supported by Grants No. FIS2008-01661, 
FIS2008-00421, FIS2008-00784, FIS 2005-03169/04627 and QOIT 
from MEC/MINCIN,  ESF/MEC
project FERMIX (FIS2007-29996-E), EU Integrated Project  SCALA, EU
STREP project NAMEQUAM, ERC Advanced Grant QUAGATUA, and Alexander
von Humboldt Foundation.

\end{document}